\documentclass[runningheads]{llncs}
\usepackage[T1]{fontenc}
\usepackage{graphicx}
\usepackage{enumitem}
\usepackage{cite} % Organizează citările (ex: [2,3] în loc de [2],[3])
\usepackage[colorlinks=true, allcolors=blue]{hyperref} 
\begin{document}
\title{RoMathExam: A Longitudinal Dataset of Romanian Math Exams (1895–2025) with a Seven-Decade Core (1957–2025)}
%\title{RoMathExam: A Longitudinal Dataset of Romanian High-School Math Exams with Topic Tags, Vectorial Embeddings Similarity, and Token-Based Solution Complexity}
% RoMathExam: 130 Years of Romanian Mathematics Exams for Curriculum Analytics, Retrieval, and Difficulty Modeling
%
\titlerunning{RoMathExam: Romanian Math Exam Dataset}
\author{Luca-Nicolae Cuclea\inst{1}\orcidID{0009-0006-6195-1354} \and
Sabin-Codruț Badea\inst{1}\orcidID{0009-0004-0510-9184} \and
Adrian-Marius Dumitran\inst{1,2,3}\orcidID{0009-0005-3547-5772}}

\authorrunning{L.-N. Cuclea et al.}

\institute{University of Bucharest, Faculty of Mathematics and Computer Science, Bucharest, Romania \\
\email{\{luca-nicolae.cuclea, sabin-codrut.badea, marius.dumitran\}@unibuc.ro} \and
Cu Drag si Sport SRL, Bucharest, Romania \and Softbinator Technologies, Bucharest, Romania}
\maketitle
%\author{Luca-Nicolae Cuclea\inst{1} \and
%Sabin-Codruț Badea\inst{1} \and
%Adrian-Marius Dumitran\inst{1}}
%
%\authorrunning{S.-C. Badea et al.}
%
%\institute{University of Bucharest, Faculty of Mathematics and Computer Science, Bucharest, Romania\\
%\email{sabin-codrut.badea@s.unibuc.ro} 
%\email{luca-nicolae.cuclea@s.unibuc.ro} 
%\email{marius.dumitran@unibuc.ro}}
%\maketitle              % typeset the header of the contribution
%
\begin{abstract}
AI in Education research increasingly relies on authentic, curriculum-grounded assessment data, yet large, well-structured exam corpora remain scarce for many languages and educational systems. We introduce RoMathExam, a longitudinal dataset of Romanian high-school mathematics exams spanning 1895--2025, with a robust standardized core for 1957--2025. The dataset contains 10,592 mathematics problems organized into 600+ complete exam sets across multiple tracks (M1--M4), covering both official national examination sessions and ministry-published training variants.

Beyond high-fidelity digitization and a unified JSON schema with traceable provenance, RoMathExam is enriched with curriculum-aligned topic tags and dense text embeddings, enabling variant detection, deduplication, and similarity-based retrieval. To overcome the lack of historical psychometric data, we propose and validate a solution complexity metric as a scalable intrinsic proxy for difficulty. Our evaluation across three frontier reasoning models (GPT-5-mini, DeepSeek-R1, and Qwen3-235B-Thinking) reveals high cross-model synchronization ($r > 0.72$), confirming the metric's ability to isolate intrinsic mathematical depth from stochastic generation noise. We demonstrate the dataset's utility through a longitudinal analysis that quantifies a ``regime shift'' from volatile historical formats to a standardized, algebra-dominant modern curriculum. RoMathExam provides a foundation for reproducible research in difficulty modeling, curriculum analytics, and LLM evaluation in low-resource linguistic contexts.

\keywords{AI in Education \and Assessment Datasets \and Mathematics Education \and Exam Analytics \and Topic Tagging \and Difficulty Modeling.}

\end{abstract}

\section{Introduction}
\label{sec:introduction}

The rapid advancement of Large Language Models (LLMs) in mathematical reasoning has exposed a critical vulnerability in evaluation frameworks: the reliance on static, potentially contaminated benchmarks. To truly measure a model's ability to generalize, researchers must look toward authentic, curriculum-grounded assessment data that mirrors the longitudinal and structured nature of human education.

However, for many national contexts, high-stakes examination data remains trapped in heterogeneous, unstructured formats across decades of shifting curricula. The Romanian educational landscape is a prime example, characterized by a rigorous national examination system and significant structural variations across academic tracks \cite{ref_oecd}. While recent efforts have improved digital access, a unified, longitudinal repository that enables cross-era comparison remains a critical gap.

We present RoMathExam, a longitudinal dataset of Romanian high-school mathematics examinations designed to support research on assessment analytics, curriculum evolution, and educational pipelines for tagging-based retrieval. The dataset contains 10,592 distinct problems organized into 600+ complete exam sets across multiple tracks (M1--M4), spanning 1895--2025 with a robust seven-decade core (1957--2025). We unify official exam sessions and ministry-published training variants under a single schema, enabling analyses both within a standardized era and across longer historical regimes.

Beyond OCRization and normalization, RoMathExam is enriched with \texorpdfstring{\\}{} curriculum-aligned topic tags and embeddings based  similarity signals. Furthermore, to overcome the lack of historical psychometric data, we propose and validate a token-based solution complexity metric as a scalable intrinsic proxy for difficulty. In summary, our contributions are:

\begin{itemize}
    \item \textbf{Longitudinal Dataset}: A high-resolution corpus of 10,592 Romanian mathematics problems spanning 130 years (1895--2025) with traceable provenance.
    \item \textbf{Curriculum Alignment}: A taxonomy of 70 curriculum-aligned topic tags and validation analyses enabling fine-grained trend studies.
    \item \textbf{Similarity Layer}: An embedding-based framework supporting deduplication, variant identification, and retrieval-based educational applications.
    \item \textbf{Complexity Proxy}: Quantified complexity according to the output token count of a Chain-of-thought (COT) LLM, with fluctuations validated by two other CoT LLMs.
\end{itemize}

\section{Background and Related Work}
\label{sec:related}
\subsection{Large-Scale Benchmarks and National Exam Datasets}
The evaluation of mathematical reasoning in Large Language Models (LLMs) has traditionally relied on static benchmarks such as GSM8K \cite{ref_46} and MATH \cite{ref_50}. However, recent work suggests that headline scores can be fragile and susceptible to dataset contamination, which requires more robust evaluation protocols \cite{ref_142}.\cite{ref_careful} and the introduction of the GSM-Symbolic benchmark by \cite{ref_mirzadeh2024gsmsymbolic}  \cite{ref_mirzadeh2024gsmsymbolic} further demonstrate that model performance is often sensitive to minor variations in problem phrasing or numerical constants.

A complementary direction is to evaluate models on authentic, curriculum-grounded assessment items drawn from high-stakes national examinations. Prior work has leveraged Brazil's ENEM\cite{ref_1, ref_23} to model item properties directly from question text, and the Chinese Gaokao\cite{ref_31} has been used as a large-scale, multi-subject benchmark for LLM evaluation. Despite these advances, structured exam repositories remain scarce for many lower-resource languages and educational systems \cite{ref_772,ref_805}.

In the specific context of Romanian resources, \cite{ref_romath} recently introduced a dataset focused on standardizing contemporary mathematics problems for model evaluation. Complementing this effort, the corpus presented here prioritizes longitudinal continuity and metadata resolution to support curriculum-aware analytics. Rather than broad categorical labels, content is organized using a detailed taxonomy of 70 topic tags and precise provenance metadata—including year, session, and difficulty tracks (M1--M4). This structural depth, spanning from 1895 to 2025, is designed to facilitate historical trend analysis and fine-grained difficulty modeling alongside standard performance benchmarking.

\subsection {Representation of Learning Domains} A core task in educational data mining is the association of assessment items with Knowledge Components (KCs) \cite{ref_2}. While recent advancements focus on automated mapping via LLMs \cite{ref_92,ref_687}, a persistent challenge remains in identifying "isomorphic" problems—items sharing underlying solution structures despite surface variations \cite{ref_138,ref_300}. While specialized benchmarks like AlgoSimBench \cite{ref_699} identify these via solution-step traces, this approach is often inapplicable to historical archives lacking step-by-step gold standards. RoMathExam addresses this by providing curriculum-aligned tags that allow for the study of semantic similarity in relation to textual embeddings. This bridges the gap between surface-level Natural Language Processing (NLP) features and deep conceptual isomorphism, a limitation noted in prior cross-lingual studies \cite{ref_603,ref_772}.

\subsection{Item Complexity and Difficulty Estimation}
In the absence of student response logs, item complexity serves as a measurable proxy for difficulty. \cite{ref_59} and \cite{ref_746} define complexity as an inherent structural property—governed by solution length and conceptual density—distinguished from difficulty, which is an observed performance measure. Recent studies confirm that LLM-based features, including token counts and solution steps, correlate strongly with human-rated difficulty parameters \cite{ref_375,ref_23}. Following this framework, RoMathExam utilizes token-based solution complexity to estimate relative difficulty across 130 years of exams.

\section{Dataset}

\subsection{Overview and Statistics}
\label{sec:stats}
The RoMathExam dataset comprises \textbf{10,592 distinct problems}. Regarding metadata completeness, marking schemes are available for 4,333 items, while specific session information is preserved for 8,445 items. The content is structured using a taxonomy of 70 unique topic tags.

A defining feature of this corpus is its temporal depth. While other recent contributions to the field focus primarily on the post-2007 digital era (e.g., \cite{ref_romath}), RoMathExam includes a significant subset of 2,320 items from the pre-2007 period. This historical coverage addresses a gap in current benchmarks, enabling the analysis of long-term curriculum evolution beyond the scope of contemporary testing standards. 

Our dataset si also available on Zenodo \cite{romathexam_zenodo}.

\subsection{Data Standardization and Profile Mapping}
\label{sec:standardization}
The dataset is distributed in JSON format via the HuggingFace Hub repository \texttt{RoMathExam/RoMathExam}\footnote{\url{https://huggingface.co/datasets/RoMathExam/RoMathExam}}, with the full schema and enumeration values detailed in the README.

To facilitate longitudinal analysis across the entire timeline, we normalized historical track designations to align with the modern four-track system. Specifically, historical "Real" and intensive mathematics variants were mapped to \textbf{M1 (Mathematics-Informatics)}, while "Uman" and general science variants were aligned with \textbf{M2 (Natural Sciences)}.

\subsection{Collection and Parsing Pipeline}
\label{sec:pipeline}

The raw documents were sourced from established digital archives of the Romanian mathematical community, primarily the historical collections curated by Sorin Borodi\cite{ref_borodi} and the standardized modern variant repositories of Pro-Matematica\cite{ref_promat}.To facilitate content extraction, the entire corpus underwent optical character recognition (OCR) and conversion using Mathpix, transforming the source PDF content into Markdown format with embedded \LaTeX\ equations.

A two-stage pipeline utilizing GPT-5-mini (via OpenAI Batch API) was employed to structure this raw text. In the first stage, the document content was processed alongside its relative file path. This path information served as a contextual anchor, allowing for the inference of information often missing from historical document headers (i.e the year of the exam might not appear in the contents, but would appear in the file name). Rather than direct text extraction, the process identified unique start/end text delimiters for each exam instance. This approach also facilitated the alignment of marking schemes (\textit{bareme}); regardless of whether solutions were interleaved within the same file or located in paired documents, the specific text boundaries for both "Subject" and "Solution" blocks were identified simultaneously.

Atomic problems were programmatically sliced and parsed into the target schema, associating exercises with grading keys into intermediate JSON objects for validation. This transition was highly cost-efficient: high-fidelity OCR (Mathpix) cost \textbf{\$52.16}, while structural parsing and 1536-D embedding generation cost \textbf{\$58.00}. This \textbf{\textasciitilde\$110} total demonstrates that building century-scale datasets for regional languages is feasible under limited resource constraints,a key consideration for frugal AIED research.

\subsection{Manual Verification and Quality Assurance}
To ensure dataset fidelity, we established a verification pipeline involving a mathematics teacher and 20 \textbf{volunteers} from a 12th-grade cohort (aged 18--19, M1 track). This activity was conducted as a pedagogical exercise in data literacy, where students identified subtle semantic errors \cite{ref_50}. The workflow proceeded in three stages:

\begin{itemize}
    \item \textbf{Visualization}: Validators used a lightweight HTML tool to render raw JSON and \LaTeX\ into human-readable exam formats, mirroring the final test setting \cite{ref_46}.
    \item \textbf{Annotation}: Using structured spreadsheets, the team reviewed batched items, flagging discrepancies (e.g., missing constraints, malformed equations) with descriptive comments.
    \item \textbf{Correction}: Flagged entries were manually reviewed and resolved \textbf{exclusively by the authors} to ensure archival integrity.
\end{itemize} 

\section{Tags and Embeddings}
\label{sec:tagging}

\subsection{Taxonomy and Automated Annotation}
To enhance the dataset's utility for educational research, we developed a granular categorization schema designed to capture the specific nuances of the Romanian Baccalaureate curriculum. The taxonomy was collaboratively created with an experienced high-school mathematics teacher to prioritize specific sub-domains (e.g., \textit{"logarithmic equations"} or \textit{"matrix inversability"}) rather than broad categories (e.g., \textit{"Algebra"}). While this granularity increases annotation complexity, it allows for more precise content analysis.

Given the dataset's extensive temporal depth (1895--2025), manual annotation of the full corpus was unfeasible. Leveraging recent findings on the efficacy of LLMs for zero-shot data labeling \cite{gilardi2023}, we employed \textbf{Gemini 3.0 Flash}. The model was provided with the pre-defined taxonomy and a strict system prompt requiring JSON output selected exclusively from the provided list.

To prevent the model from assigning technically correct but non-essential tags, such as labeling a matrix determinant problem as \textit{"real number calculations"}, we enforced a constraint to ensure tags capture the \textit{essence} of the problem. The limit was defined as a maximum of 3 tags for single-item exercises, and $n+2$ tags for exercises with $n$ sub-points.

\subsection{Tagging Validation Experiments}

To validate the automated annotation, we implemented a multi-stage verification process. Initial automated checks ensured that every exercise was assigned at least one tag, that all tags were part of the defined tag list and that each tag was used a statistically relevant amount of times. 

Subsequently, we performed a blind manual annotation of 100 randomly sampled exercises by a domain expert. We calculated similarity metrics between the expert's tags and the LLM's tags. The results indicate a high degree of alignment \cite{mchugh2012}, as detailed in Table~\ref{tab:metrics}.

\begin{table}[ht]
    \centering
    \caption{Similarity metrics between Expert and LLM annotations (N=100)}
    \label{tab:metrics}
    \begin{tabular}{lc}
        \hline
        \textbf{Metric} & \textbf{Value} \\
        \hline
        Jaccard Similarity & 0.8847 \\
        Precision & 0.8811 \\
        Recall & 0.9662 \\
        F1 Score & 0.9217 \\
        Cohen's Kappa & 0.9477 \\
        \hline
    \end{tabular}
\end{table}
Qualitative analysis of the discrepancies revealed three primary disagreement types: 
\begin{itemize}
    \item \textbf{Granularity Noise}: The LLM included generic tags for minor steps (e.g., tagging trigonometric operations within a matrix problem).
    \item \textbf{Method-Dependent Tagging}: The LLM inferred specific solution paths (e.g., tagging "derivatives" for limits solvable by L'Hôpital's rule).
    \item \textbf{Ontological Ambiguity}: Arising when multiple valid strategies existed (e.g., tagging "vectors" vs. "geometry" for shape vertices).
\end{itemize}

These experiments suggest that the automated tags are highly reliable in identifying the core content of the exercises. We argue that a perfect agreement (1.0) is practically unattainable with a taxonomy of this granularity due to the inherent subjectivity of mathematical problem classification, a well-documented phenomenon in linguistic and semantic annotation tasks \cite{artstein2008}. While using broader categories (e.g., Geometry vs. Analysis) would almost certainly generate even higher agreement, it would drastically reduce the utility of the dataset.

\subsection{Textual Embeddings}
\label{sec:embeddings}

Semantic representations were generated using the \texttt{text-embedding-3-small} model, producing 1536-dimensional vectors for each item. To ensure that similarity metrics reflect only the mathematical content, the model was provided exclusively with the raw problem statements. All secondary information, including topic tags, difficulty proxies, and temporal metadata, was strictly excluded from the input to prevent any external bias in the embedding space.

\subsection{Longitudinal Curriculum Analysis}
\label{sec:longitudinal_analysis}

With the granular tagging taxonomy validated and high-dimensional embeddings generated, we utilized the annotated dataset to analyze the temporal evolution of the curriculum. Our analysis reveals that the Romanian Baccalaureate has undergone significant pedagogical "phase transitions".

\begin{figure}[ht]
    \centering
    \begin{minipage}{0.49\textwidth}
        \centering
        \includegraphics[width=\linewidth]{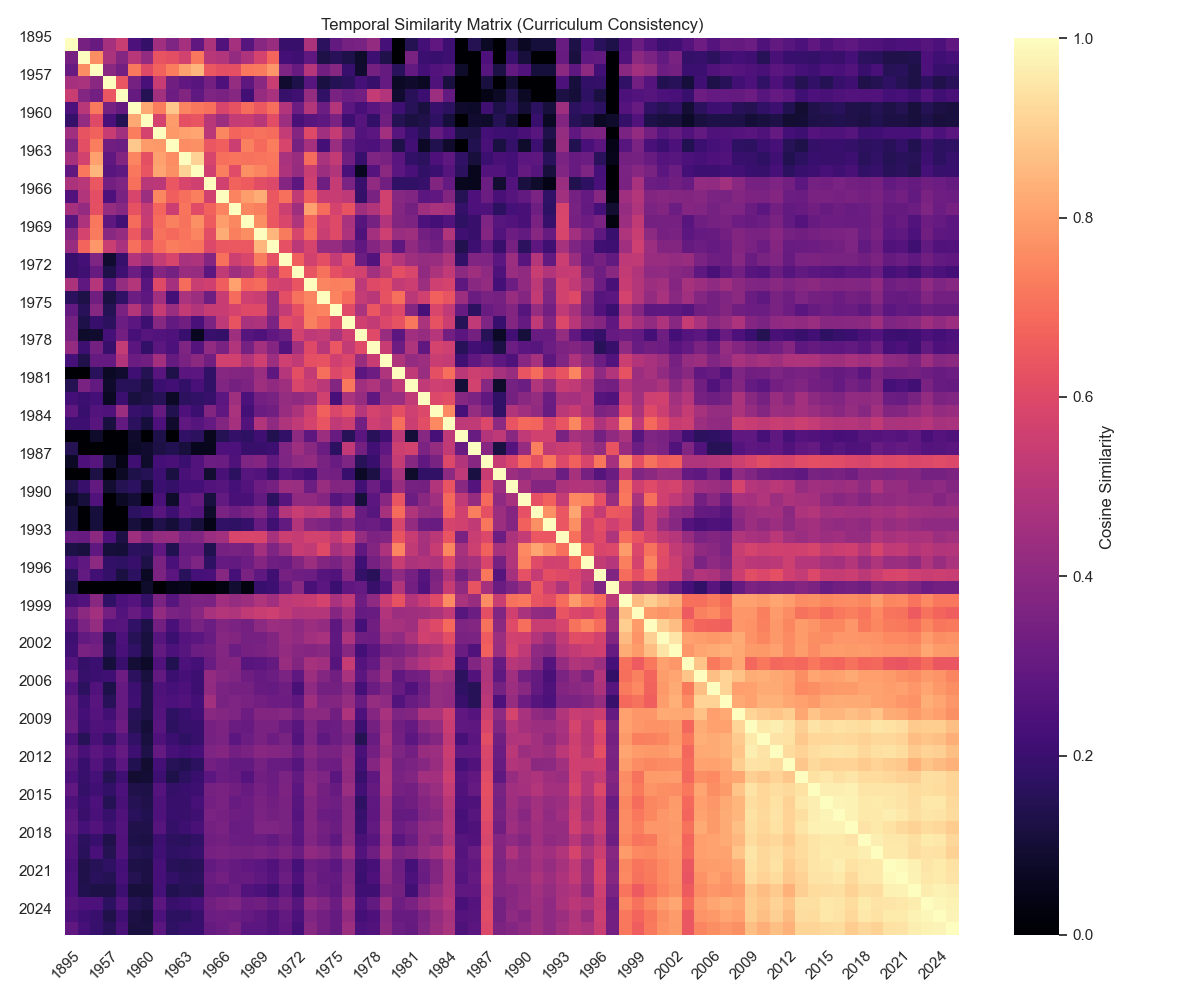}
        \small{(a) Curriculum Consistency (Tags)}
    \end{minipage}\hfill
    \begin{minipage}{0.46\textwidth}
        \centering
        \includegraphics[width=\linewidth]{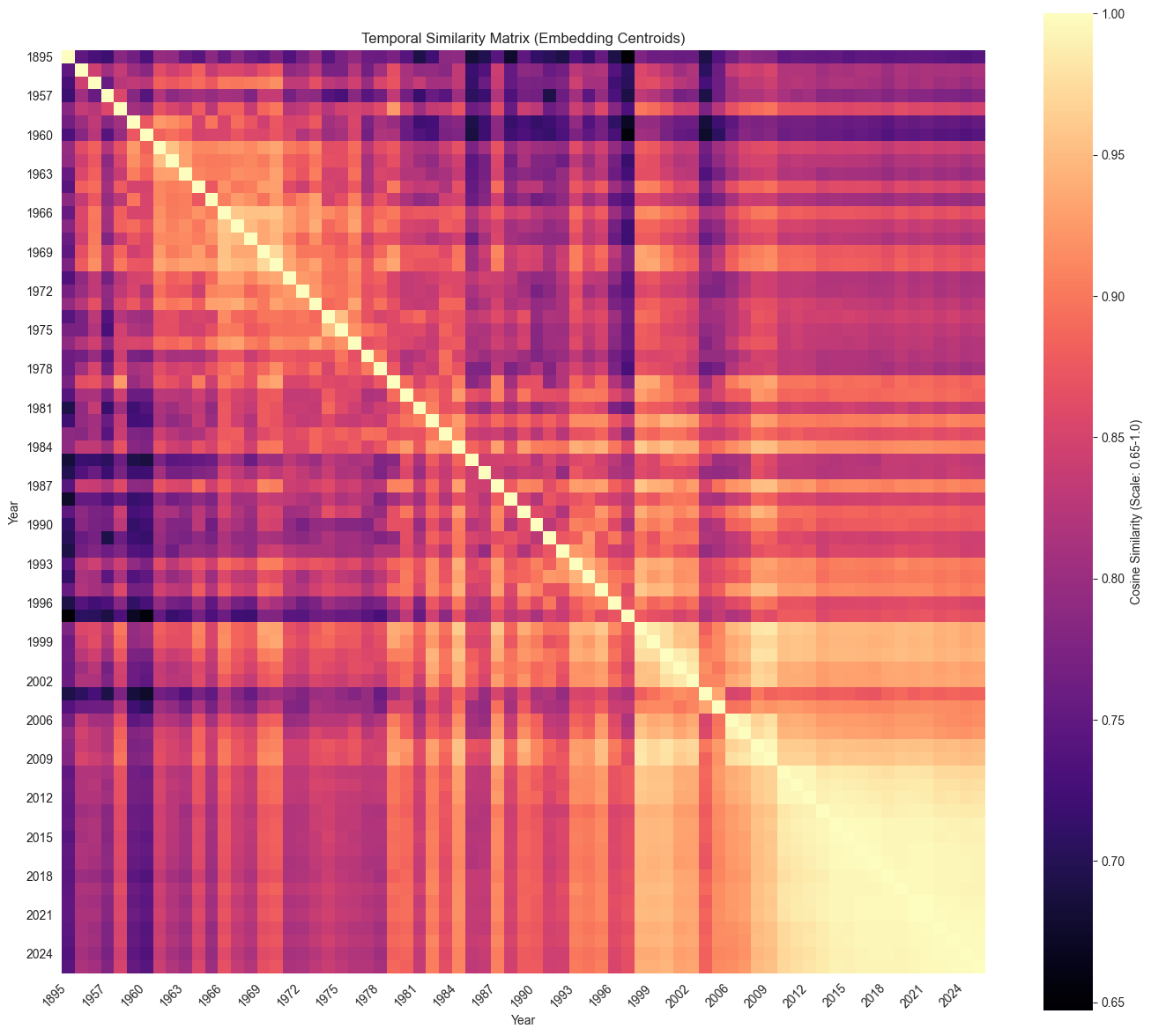}
        \small{(b) Embedding Centroids Similarity}
    \end{minipage}
    \caption{Temporal Similarity Matrices comparing year-to-year consistency.}
    \label{fig:similarity}
\end{figure}

\textbf{Curriculum Consistency and Regime Shifts.}
To quantify the stability of the examination format over time, we computed two distinct \textit{Temporal Similarity Matrices} (Figure~\ref{fig:similarity}). 

First, adopting quantitative methods from the field of culturomics \cite{michel2011}, we calculated the cosine similarity between the topic distribution vectors of every pair of years (Figure~\ref{fig:similarity}a). Second, to corroborate these findings using latent semantic information, we computed a centroid embedding for each year by averaging the vector embeddings of all exercises within that year. We then calculated the cosine similarity matrix between these yearly centroids (Figure~\ref{fig:similarity}b).

Both heatmaps independently reveal a striking antithesis between the 20th and 21st centuries. The period prior to 2000 is characterized by low cross-temporal similarity (indicated by the darker regions), suggesting a highly volatile curriculum where the structure of the exam changed radically as years passed. In contrast, the post-2000 era exhibits a distinct square structure of high similarity in both the explicit tag-space and the latent embedding-space. This indicates the emergence of a standardized examination framework where the distribution of mathematical concepts remains stable over time.

The high congruence between the explicit taxonomy-based matrix and the implicit embedding-based matrix serves as a powerful dual validation. It demonstrates that the observed "regime shift" is a genuine property of the curriculum's evolution, not an artifact of a specific annotation scheme. Furthermore, the ability of the embedding centroids to recover the same temporal structure as the tags suggests that the embeddings successfully capture the pedagogical essence of the exercises, validating their quality for downstream tasks.

\begin{figure}[ht]
    \centering
    \includegraphics[width=\linewidth]{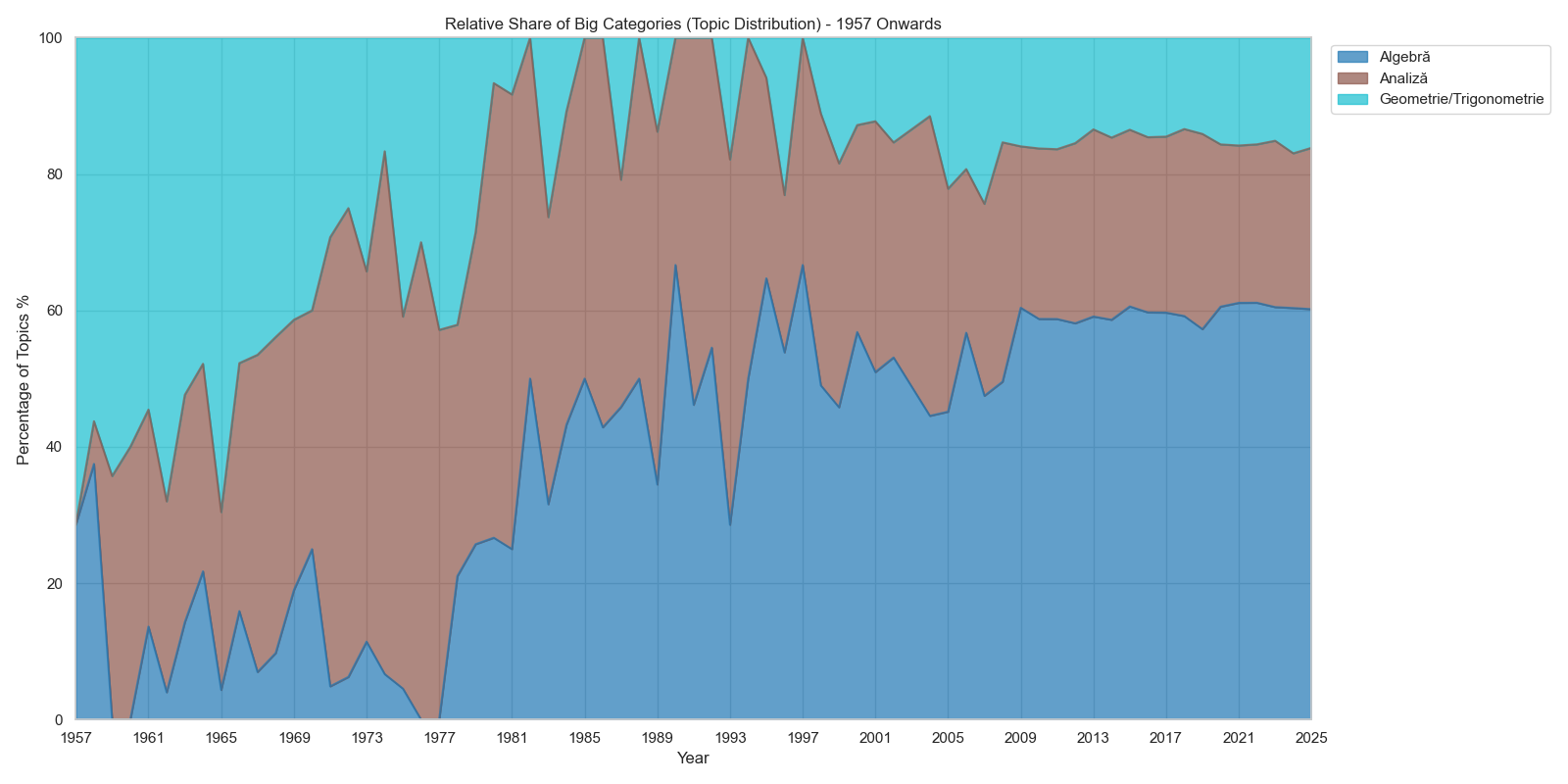}
    \caption{Relative Share of Major Mathematical Categories (1957--2025). The plot highlights the gradual dominance of Algebra in the modern curriculum.}
    \label{fig:distribution2}
\end{figure}

\textbf{Topic Distribution Drift.}
Finally, an analysis of the three major domains; Algebra, Analysis, and Geometry; confirms a long-term concept drift (Figure~\ref{fig:distribution2}). While the early 20th century placed a heavy emphasis on Geometry and Analysis, the modern era is dominated by Algebra, which now consistently accounts for approximately 60\% of the examination content. This structural evolution further reinforces the necessity of temporally-aware training splits when developing Intelligent Tutoring Systems (ITS) based on this corpus, as standard random splitting would likely overestimate model performance in such non-stationary environments \cite{sogaard2021}.

\section{Difficulty Estimation via Token-Based Solution Complexity}
\label{sec:difficulty}

\subsection{Motivation: The Limitations of Traditional Difficulty Metrics}
Establishing a consistent difficulty metric across a 130-year temporal span presents significant methodological challenges. The primary obstacle is the absence of a stable baseline: historical archives lack granular student performance data, and modern psychometric indicators are heavily confounded by external sociopolitical variables. Consequently, longitudinal comparisons based on pass rates or human perception are prone to significant error.

The reliability of pass rates as a proxy for item difficulty is compromised by major administrative shifts. \cite{ref_borcan} provides compelling evidence that the introduction of rigorous anti-corruption interventions in 2011—specifically the installation of CCTV monitoring in exam centers—resulted in a precipitous drop in pass rates unrelated to curriculum content. This dissociation implies that statistical difficulty often reflects the efficacy of proctoring protocols rather than the intrinsic mathematical complexity of the items.

Concurrently, subjective perception of difficulty is distorted by the standardization of the testing format. Qualitative analyses suggest that the shift toward standardized testing has moved the focus from creative reasoning to rote memorization, potentially lowering the effective cognitive ceiling despite the high-pressure environment \cite{ref_jesbs}. Furthermore, public discourse frequently conflates difficulty with political cycles, introducing anecdotal noise that renders human labeling unreliable.

To circumvent these sociopolitical confounders, Output Token Complexity is proposed as an objective, intrinsic proxy. By utilizing LLMs as "synthetic students," the mathematical signal is isolated from external stressors. The central hypothesis is that for a sufficiently capable reasoning model, the length of the generated solution path correlates with the logical depth and number of inference steps required, providing a sterile measure of complexity.

\subsection{Methodology and Architecture-Agnostic Validation}
To operationalize this metric, a batch processing pipeline was implemented using \texttt{gpt-5-mini} as the primary evaluator for the full dataset ($N = 10592$). To mitigate the risk of prompt-induced verbosity, a minimal system instruction was employed: \textit{"Ai rolul de a rezolva probleme de matematică"} ("Your role is to solve math problems"). This unconstrained prompt enables the model to dynamically determine the necessary depth of explanation based solely on the problem's requirements.

Validation was conducted to ensure that the resulting complexity signal is an intrinsic property of the dataset rather than an artifact of a specific model architecture. A cross-model consistency check was performed using two additional high-reasoning models, \texttt{Qwen/Qwen3-235B-Thinking} and \texttt{DeepSeek-R1-0528}, restricted to the "Summer" and "Simulation" sessions of the standardization era (2010--2025).

\begin{figure}[ht]
    \centering
    \includegraphics[width=0.8\linewidth]{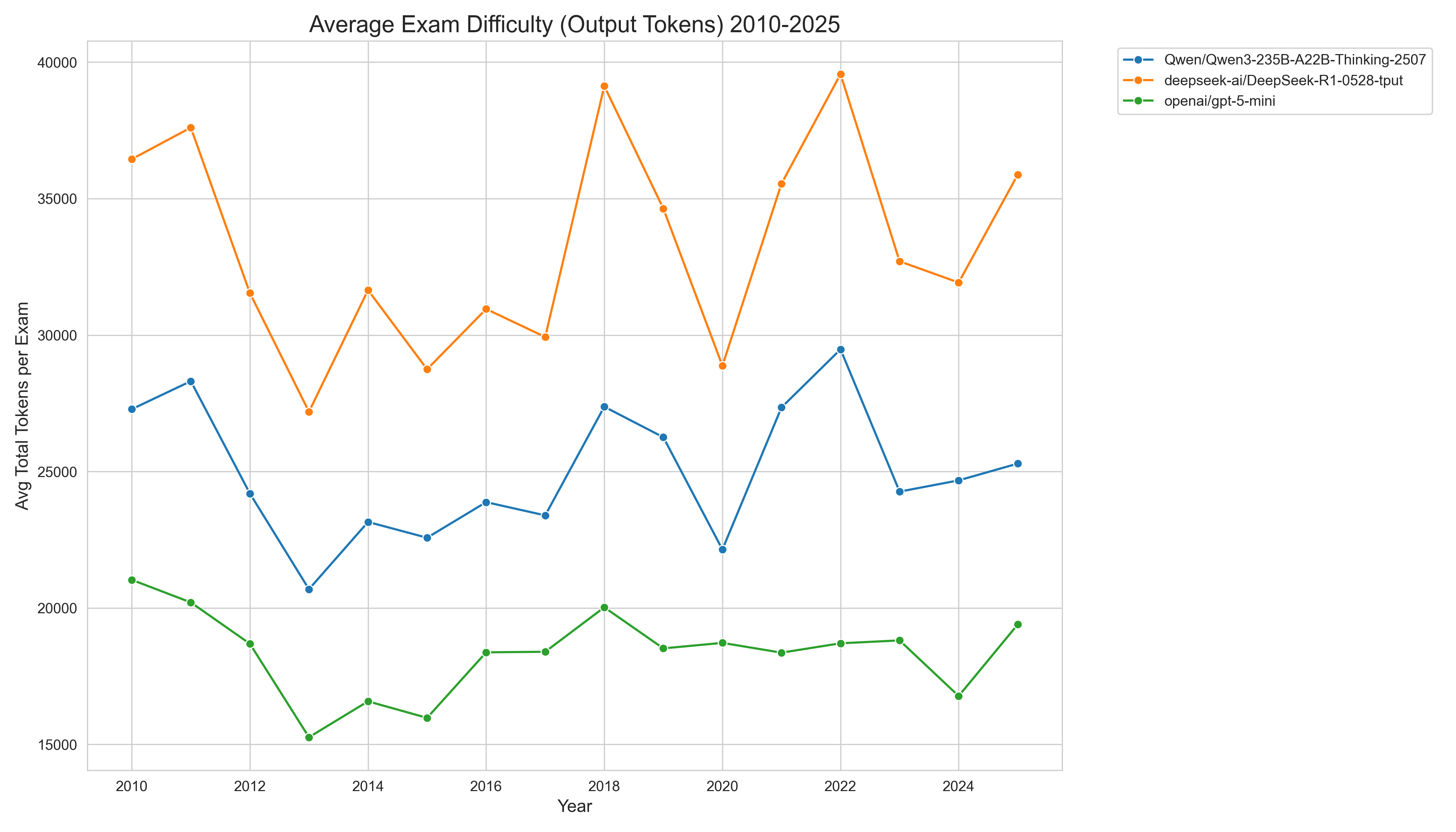}
    \caption{Cross-model complexity trends (2010--2025). High synchronization ($r > 0.72$) confirms intrinsic difficulty spikes between the models.}
    \label{fig:fluctuations}
\end{figure}

The analysis, visualized in Figure~\ref{fig:fluctuations}, reveals a high synchronization between models ($r > 0.72$). Although absolute token counts vary due to tokenizer distinctiveness, the relative longitudinal fluctuations remain consistent across all three architectures. Validating this proxy via multi-model reasoning chains cost approximately \textbf{\$93.80}, confirming that large-scale difficulty estimation is computationally accessible.
%Intra-model consistency was assessed through duplicate execution cycles, yielding a divergence of less than $28\%$, which confirms that the detected difficulty spikes are robust to stochastic generation noise.
% Aici poate trebuia in graifc puse ambele puncte....

\subsection{Findings: Structural Evolution and Syllabus Dynamics}
Applying the token-complexity metric confirms that National Simulations are consistently more difficult than final Summer sessions. Across the 2010--2025 interval, simulations yielded statistically higher token counts, supporting the hypothesis that diagnostic tests prioritize complex, integrative problem-solving to incentivize study.

\begin{figure}[ht]
    \centering
    % Left Image
    \begin{minipage}[b]{0.48\textwidth}
        \centering
        \includegraphics[width=\linewidth]{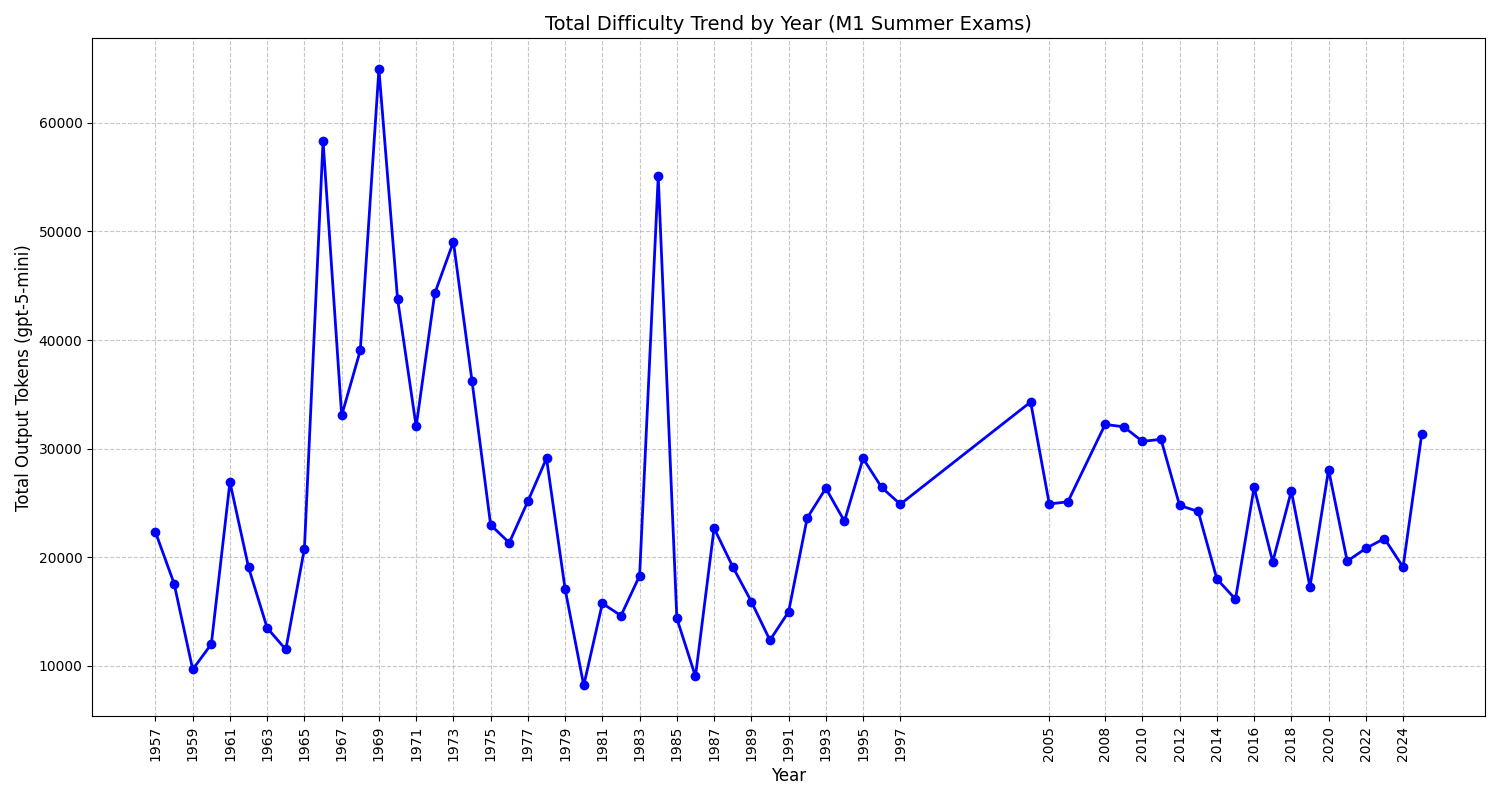}
        % Manually formatted caption (a)
        \par\vspace{2pt}
        {\small (a) Aggregate Token Volume (Exam Load)}
        \label{fig:total_tokens}
    \end{minipage}
    \hfill
    % Right Image
    \begin{minipage}[b]{0.48\textwidth}
        \centering
        \includegraphics[width=\linewidth]{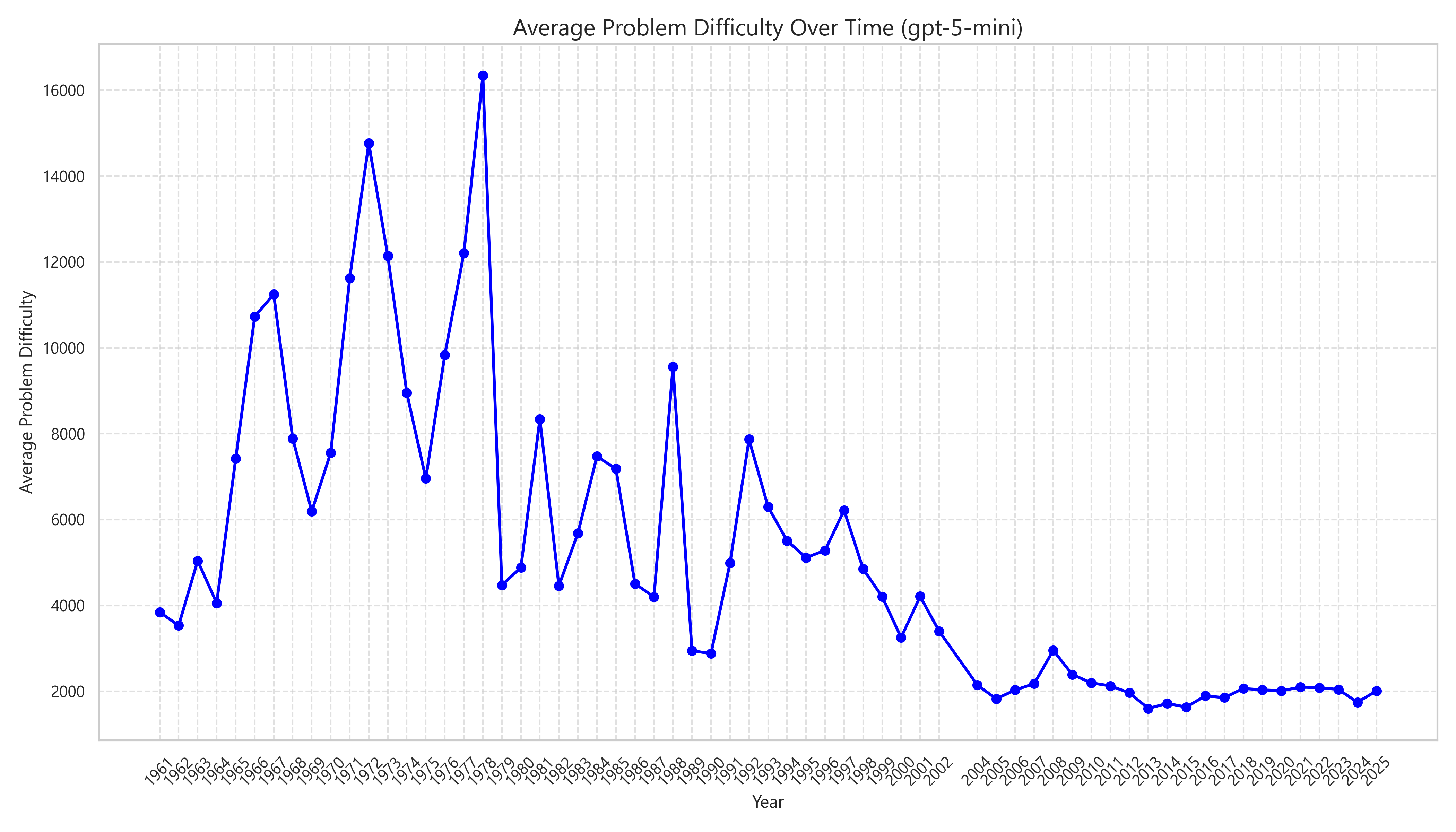}
        % Manually formatted caption (b)
        \par\vspace{2pt}
        {\small (b) Mean Tokens per Problem (Item Complexity)}
        \label{fig:avg_tokens}
    \end{minipage}
    
    \caption{The Stabilization of Difficulty Metrics. Historical exams (pre-2000) display significant stochastic variance (a), often driven by isolated, high-complexity outliers. The modern standardized era (2010--2025) demonstrates a convergence toward stability: while individual item complexity has decreased (b), the aggregate cognitive load is maintained through a higher volume of items, resulting in a consistent total output.}
    \label{fig:token_metrics}
\end{figure}

Longitudinally, Figure~\ref{fig:token_metrics} reveals a shift in the definition of difficulty. Historical exams (1950--2000) display high variance driven by idiosyncratic, high-complexity outliers. In contrast, the modern era (2010--2025) exhibits a ``Low Complexity / High Consistency'' profile: while mean per-item complexity has decreased, the aggregate token volume per exam has converged into a stable band. This reflects a pedagogical transition from difficulty based on creative derivation to one requiring sustained cognitive stamina across a standardized algorithmic load.

\section{Educational Use Cases}
\label{sec:use_cases}

\subsection{Personalized Practice and Isomorphic Retrieval}
\label{subsec:student_tools}

For students, the primary challenge in exam preparation is often not the lack of material, but the inability to locate specific practice problems that target their individual weaknesses. Traditional keyword searches are often too broad. RoMathExam addresses this by enabling \textit{embedding-based retrieval}, allowing students to perform ``isomorphic practice.''

By utilizing the semantic similarity layer (Section~\ref{sec:embeddings}), a student can input a specific problem they found difficult and instantly retrieve the $k$-nearest neighbors from the historical archive. As shown in Table~\ref{tab:retrieval_examples}, this process identifies problems that share the same underlying logical structure or solution strategy, even if the numerical values or surface phrasing differ. This allows for targeted reinforcement of specific algorithmic techniques (e.g., irrational equations involving radicals) rather than generic topic review.

\begin{table}[ht]
    \centering
    \caption{Examples of Isomorphic Retrieval. The system identifies historical variants that share the same underlying logic as the query problem, enabling targeted student practice.}
    \label{tab:retrieval_examples}
    \setlength{\tabcolsep}{6pt}
    \renewcommand{\arraystretch}{1.5}
    \begin{tabular}{|p{0.45\textwidth}|p{0.45\textwidth}|}
        \hline
        \textbf{Query Problem (Student Input)} & \textbf{Retrieved Isomorphic Variant} \\
        \hline
        % Row 1: 2025 vs 2011 (Radical Equations)
        \textit{2025, Special Session, M2:} \newline
        Rezolvați în mulțimea numerelor reale ecuația $\sqrt{2-x}=\sqrt{x^{2}+x-1}$.
        & 
        \textit{2011, Model, M2:} \newline
        Rezolvați în mulțimea numerelor reale ecuația $2-\sqrt{2-x}=x$.
        \\
        \hline
        % Row 2: 2023 vs 2010 (Geometry - Distance to Midpoint)
        \textit{2023, Special Session, M2:} \newline
        În reperul cartezian $xOy$ se consideră punctele $A(1,-5)$ și $B(5,5)$. Determinați distanța de la punctul $O$ la mijlocul segmentului $AB$.
        & 
        \textit{2010, Model, M2:} \newline
        În reperul cartezian $xOy$ se consideră punctele $A(-1,-2),\ B(1,2)$ şi $C(2,-1)$. Calculați distanța de la punctul $C$ la mijlocul segmentului $AB$.
        \\
        \hline
    \end{tabular}
\end{table}

\subsection{Stylistic Alignment for Automated Item Generation}
\label{subsec:style_alignment}
A significant challenge in utilizing LLMs for automated item generation is the "linguistic alignment gap": generic models often produce mathematically correct but stylistically unnatural phrasings that deviate from established national standards. RoMathExam serves as a high-fidelity reference corpus for \textit{Retrieval-Augmented Stylistic Alignment}. By retrieving existing problem statements that are structurally isomorphic to a draft input (e.g., a raw equation or abstract concept), an LLM can be prompted to contextualize abstract mathematical structures within the canonical linguistic register of the Romanian curriculum. This process ensures that generated items adhere to the specific syntactic conventions and imperative formulations (e.g., "Se consideră...", "Să se demonstreze că...") characteristic of the historical archive, thereby automating the production of exam-ready content \cite{ref_style_transfer}.

\section{Ethics and Limitations}
\label{sec:ethics}

\setlength{\parindent}{0pt}
\setlength{\parskip}{4pt} % Adjust this for a tiny bit of vertical air between blocks

\textbf{Provenance.} \textsc{RoMathExam} is compiled from official Romanian high-school mathematics exam materials and ministry-published training variants released for public use. 

\textbf{Copyright and Attribution.} We do not claim copyright over original problem statements. The contribution lies in the OCRization, normalization, structured metadata, curriculum-aligned tags, and derived complexity signals. 

\textbf{Privacy and Responsible Use.} The dataset contains no personal identifying information. We position RoMathExam for research and formative learning and discourage use for real-time assistance in proctored settings.

\textbf{Derived Difficulty Proxy.} Token-based complexity is an intrinsic, protocol-dependent proxy. It is not a human psychometric difficulty estimate and may vary based on model architecture and generation settings.

% Example references you can include in the bibliography:
% \bibitem{legea8_1996}
% Parlamentul Rom\^aniei: Legea nr. 8/1996 privind dreptul de autor \c{s}i drepturile conexe (art. 9).
%
% \bibitem{borodi_archive}
% Borodi, S.: Subiecte examene (arhiv\u{a} public\u{a} de subiecte). \url{https://sorinborodi.ro/subiecte_examene.html}, last accessed YYYY/MM/DD

\section{Conclusion and Future Work}
\label{sec:conclusion}

\textbf{Conclusion.} In this work, we introduced RoMathExam, a longitudinal dataset of Romanian high-school mathematics exams spanning 1895–2025. By providing a unified JSON schema, curriculum-aligned topic tags, and a cross-LLM validated token-based complexity proxy, we address the scarcity of structured, long-term assessment data for regional languages. Our analysis of regime shifts in the curriculum highlights the dataset's utility for understanding the evolution of national pedagogical standards.

\smallskip
\noindent
\textbf{Future Work.} To extend this research, we propose the following directions:
\begin{itemize}[leftmargin=*,noitemsep,topsep=2pt]
    \item \textbf{Structural Complexity:} Augmenting the complexity metric with sub-question dependency ratios and step-to-answer estimates.
    \item \textbf{Multimodal Expansion:} Developing machine-readable diagram representations for geometry-heavy Olympiad subsets.
    \item \textbf{Multilingual Support:} Releasing an English-translated version to facilitate global cross-national pedagogical benchmarking.
\end{itemize}

\smallskip
\noindent
\textbf{Reproducibility and Data Access.} To support the AIED community, we release:
\begin{itemize}[leftmargin=*,noitemsep,topsep=2pt]
    \item \textbf{Dataset:} The full RoMathExam corpus, metadata, and 70-topic tags [Link].
    \item \textbf{Embeddings:} Precomputed 1536-D semantic vectors to minimize community costs.
    \item \textbf{Pipeline:} Python scripts for parsing, complexity calculation, and synchronization.
\end{itemize}

\subsubsection*{Acknowledgments}
We would like to thank Together AI for providing the API credits used for model evaluation. 
Additionally, the third author acknowledges the support provided by Softbinator Technologies 
in fostering educational and research initiatives within the University of Bucharest.

\end{document}